\pdfoutput=1
\documentclass[notitlepage,twocolumn,superscriptaddress, aps, prl, floatfix]{revtex4-1}
\usepackage{lmodern}
\usepackage{textcomp}
\usepackage{graphicx}
\usepackage{amsmath}
\usepackage{mathtools}
\usepackage{graphicx}
\usepackage{bm}
\usepackage{array}
\usepackage{dcolumn}
\usepackage{hepparticles}
\usepackage{heppennames}
\usepackage{hepnicenames}
\usepackage[colorlinks]{hyperref}
\usepackage{lineno}
\usepackage[english]{babel}
\usepackage[autostyle]{csquotes}
\usepackage{physics}
\usepackage{qcircuit}
\newcommand{\Figblackhole}{
\begin{figure}
\centering
\includegraphics[width = 89 mm]{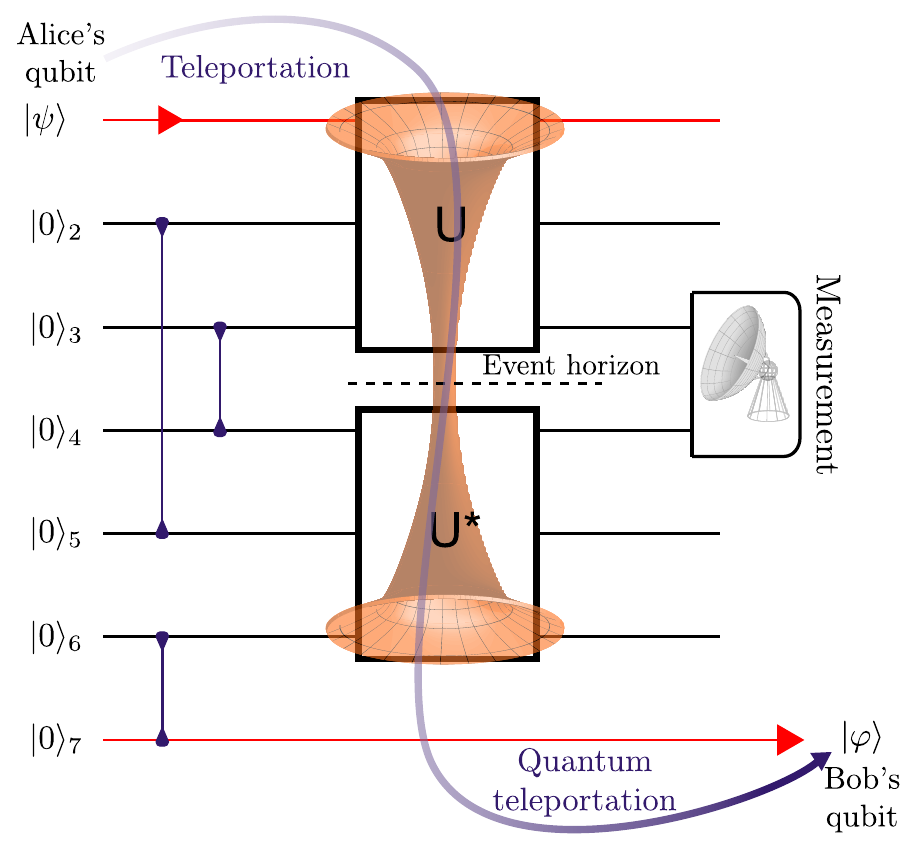}
\caption{Schematic of our 7-qubit scrambling circuit with Alice's qubit, $\ket{\psi}$, as an input. Alice's information is scrambled throughout the entire system by maximally scrambling Clifford circuits $\hat{U}$ and $\hat{U}^*$. If the scrambling operations are performed without errors, Bob will be able to teleport Alice's qubit to himself $(\braket{\varphi}{\psi} = 1)$ by  a projective measurement. The underlay depicts an interpretation of the unitaries $\hat{U}$ and $\hat{U}^*$ as representing a two-sided black hole connected by a wormhole \cite{gao2017traversable, maldacena2017diving, SusskindWormhole2017}; in the experiment, the two-sided black hole is modeled via multiple EPR pairs as in \cite{maldacena2013cool}. The maximally scrambling nature of the black hole dynamics is captured by $\hat{U}$ and $\hat{U}^*$. The vertical lines indicate that qubit pairs $\{3,4\}$, $\{2,5\}$, and $\{6,7\}$ are initially prepared as EPR pairs, $\ket{\text{EPR}} = \frac{1}{\sqrt{2}} (\ket{00}+\ket{11}$).}
\label{Fig1}
\end{figure}
}

\newcommand{\Figscanscramble}{
\begin{figure*}
\begin{center}
\includegraphics[width = 180 mm]{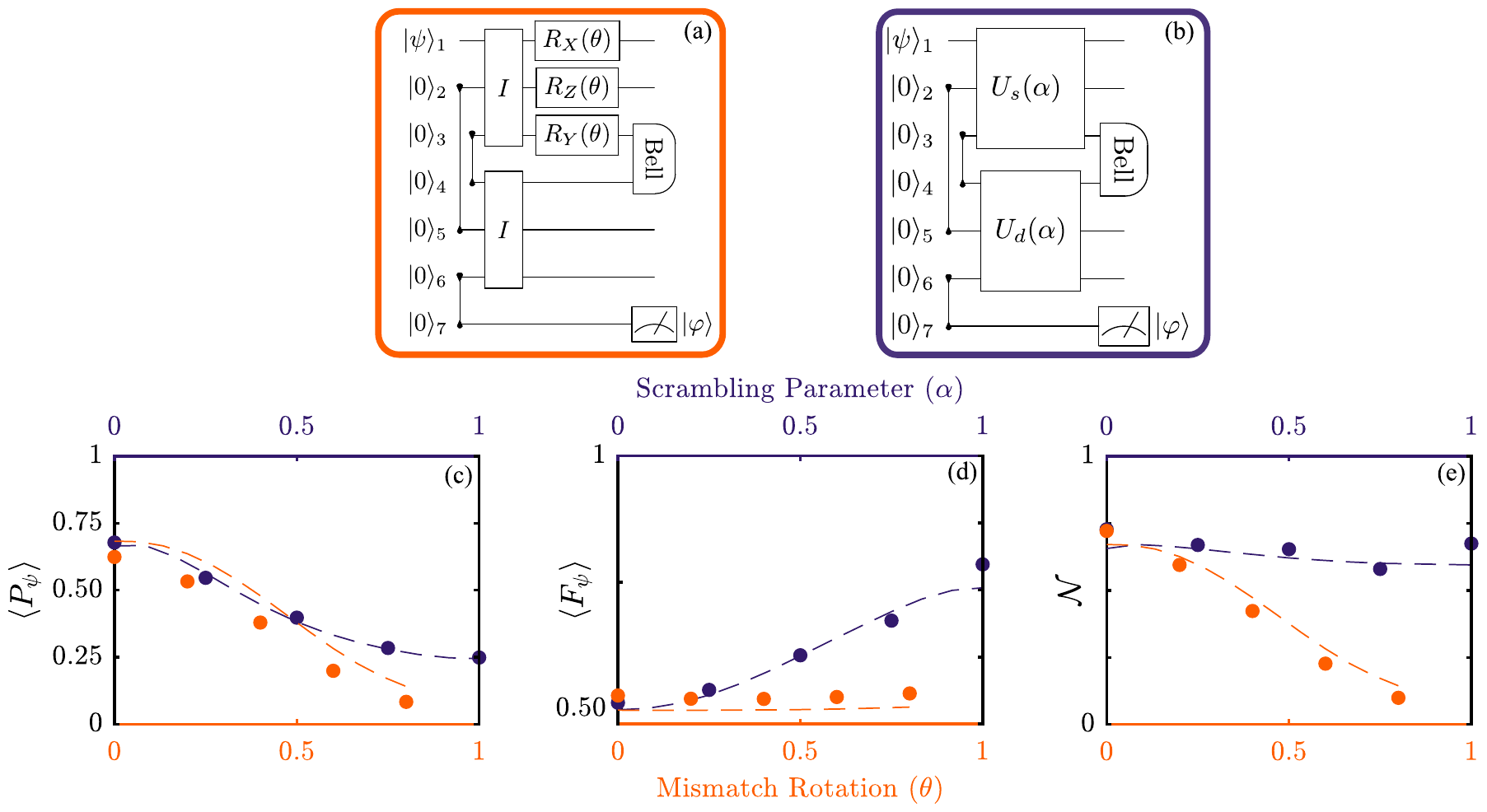}
\end{center}
\caption{(a) Circuit designed to demonstrate that a mismatch between $\hat{U}_s$ and $\hat{U}_d$ naturally leads to the decay of the OTOC without enabling teleportation. Following the $\hat{U}_s = I$ operation, we perform three additional  independent rotations $R_X$, $R_Y$, and $R_Z$ on the qubits by angle $\theta$.  Accompanying data (orange) for the averaged successful projective measurement ($\langle P_\psi \rangle$), averaged teleporation fidelity ($\langle F_\psi \rangle$) and noise factor ($\mathcal{N}$) as a function of $\theta$ are depicted in panels (c-e). (b) Circuit designed to probe the OTOC and teleportation fidelity as a function of the scrambling parameter $\alpha$ with $\alpha = 0$ representing no scrambling and  $\alpha = 1$ representing full scrambling. Accompanying data (purple) for $\langle P_\psi \rangle$, $\langle F_\psi \rangle$ and $\mathcal{N}$  as a function of $\theta$ are depicted in panels (c-e). (c-e) For the mismatch circuit shown in (a), we find that the teleportation fidelity remains near its minimal value, $\langle F_\psi \rangle \sim 0.5$, for all $\theta$, consistent with our expectation that scrambling is not occurring. However, one observes that the OTOC (as measured via $\langle P_\psi \rangle$)  decays to nearly zero, which would nominally suggest scrambling. This is precisely the challenge with interpreting OTOC measurements as an indicator of scrambling in noisy experiments. Finally, as expected, we observe that the noise parameter $\mathcal{N}$ decays as the mismatch grows. For the tunable scrambling circuit shown in (b), we find that the teleportation fidelity increases as we increase our scrambling parameter $\alpha$. This increase in teleportation fidelity is accompanied by a decrease in the OTOC (as measured via $\langle P_\psi \rangle$), indicating that the OTOC's decay is caused, at least in part, by true scrambling dynamics.  Finally, the noise parameter $\mathcal{N}$ remains relatively constant because the complexity and therefore experimental errors associated with implementing $\hat{U}(\alpha)$ are mostly  $\alpha$-independent. Dashed lines represent theory curves that are obtained via numerical simulations of the circuit assuming a one-parameter coherent error model (see Appendix for details). Error bars indicating statistical uncertainties are smaller than the data points and are omitted.}
\label{ScanScramble}
\end{figure*}
}

\newcommand{\Teleportation}{
\begin{figure*}
\centering
\includegraphics[width = 180 mm]{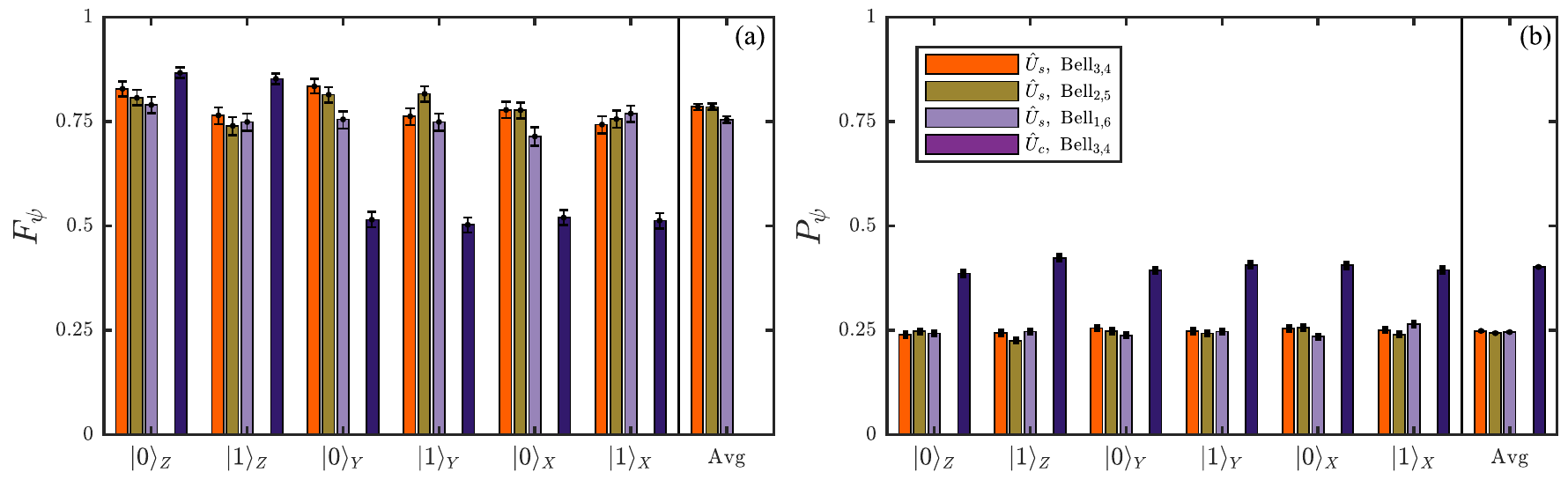}
\caption{
(a) Teleportation fidelities ($F_{\psi}$) for maximally scrambling ($\hat{U}_s(\alpha = 1)$) and classically scrambling ($\hat{U}_c$) unitaries are presented for all teleported states as well as all subsystems $\{3,4\}$,  $\{2,5\}$, and  $\{1,6\}$ that was used for the projective measurement (indicated as different bar colors). In the case of the maximally scrambling unitary, all basis states and all measurement Bell pairs lead to successful teleportation, demonstrating the full delocalization of Alice's quantum state. In the case of the classical scrambling unitary, we projectively measure on subsystem $\{3,4\}$. Only the z-basis states are successfully teleported. Data (for $\hat{U}_s$) averaged over all six teleported states is shown in the final column. Data is depicted with the same color scheme as in (b).  b) Measurements of $P_{\psi}$ from the experiments described in (a). The probabilities averaged over all basis states constitutes the average experimental OTOC. Error bars indicate statistical uncertainties.
}
\label{FigScramble}
\end{figure*}
}

\newcommand{\Grover}{
\begin{figure}
\begin{center}
\includegraphics[width = 90 mm]{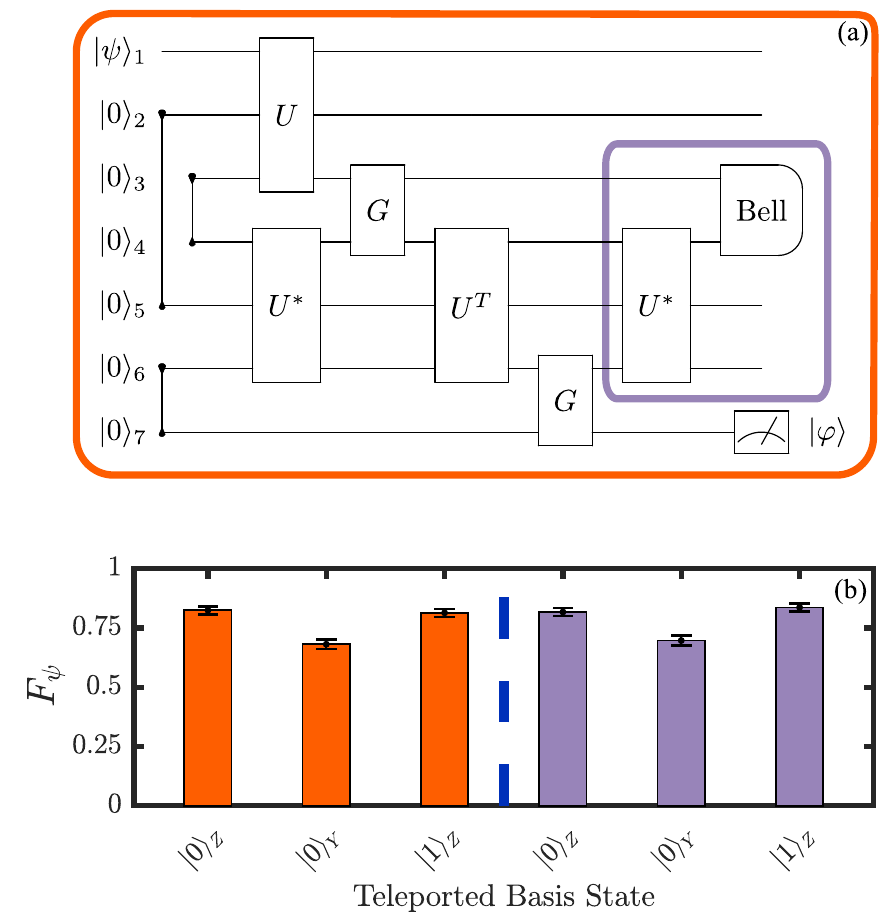}
\end{center}
\caption{(a) Circuit diagram for our deterministic teleportation scheme that utilizes a built in Grover's search protocol. Ideally, this search finds the desired EPR state with perfect fidelity; by adding in post-selection (purple box), we can quantify the performance of the search. (b) Depicts the measured  teleportation fidelity for different initial states utilizing the Grover search protocol, both with and without post-selection. The average fidelity  without post-selection (orange) is $77(2)\%$ and with post-selection (purple) is $78(2)\%$.}
\label{FigOther}
\end{figure}
}
\newcommand{\AppendixA}{
\subsection{Appendix A: Experimental Details}
\subsubsection{Trapped Ion Qubits}
We perform the experiment on a quantum computer consisting of a chain of nine $^{171}$Yb$^+$ ions confined in a Paul trap and laser cooled near the motional ground state. The hyperfine-split $^2S_{1/2}$ ground level with an energy difference of $12.642821\:$GHz provides a pair of qubit states, $|0\rangle=|0,0\rangle$ and $|1\rangle=|1,0\rangle$ with quantum numbers $|F,m_F\rangle$, that are magnetic field independent to first order. The $1/$e-coherence time of this so-called ``atomic clock'' qubit is $1.5(5)\:$s in our system, limited by magnetic field noise. Optical pumping is used to initialize the state of all ions, and the final states are measured collectively via state-dependent fluorescence detection \cite{Olmschenk07}. Each ion is mapped to a distinct channel of a photomultiplier tube (PMT) array. The average state detection fidelity is $99.4(1)\%$ for a single qubit, while a $7$-qubit state is typically read out with $92(1)\%$ average fidelity, limited by channel-to-channel crosstalk. These state detection and measurement (SPAM) errors are characterized in detail by measuring the state-to-state error matrix. 

\subsubsection{Gate Operations}
Quantum operations are achieved by applying two Raman beams from a single $355\:$nm mode-locked laser, which form beat notes near the qubit frequency. The first Raman beam is a global beam applied to the entire chain, while the second is split into individual addressing beams to target each ion qubit \cite{debnath_demonstration_2016}, controlled by a set of Arbitrary Waveform Generators (AWGs). Single qubit gates are generated by driving resonant Rabi rotations (R-gates) of defined phase, amplitude, and duration. Single-qubit Z-rotations are applied efficiently as classical phase advances. Two-qubit gates (XX-gates) are realized by illuminating two ions with beat-note frequencies near the motional sidebands and creating an effective spin-spin Ising interaction via transient entanglement between the state of two ions and all modes of motion \cite{Molmer99}. To ensure that the motion is left disentangled from the qubit states at the end of the interaction, we employ a pulse shaping scheme \cite{Zhu06eur,Choi14}.  We use only the middle seven ions in the chain as qubits, in order to ensure a higher uniformity in the ion spacing, matching the equally-spaced individual addressing beams. The two edge ion qubits are neither manipulated nor measured, but their contribution to the collective motion is included when creating the entangling operations. 

\subsubsection{Bell State Preparation and Measurement}
In Fig.~\ref{FigScramble}b, we depict Bell state pairs initially in the state $\frac{1}{\sqrt{2}} \ket{00}+\ket{11}$. This entangled state is created using the following circuit:
\begin{equation}
\centering
\Qcircuit @C=1em @R=0.3em {
&   \qw	&\multigate{1}{XX(\pm \frac{\pi}{4})}&\gate{R_z(\pm\frac{\pi}{2})}&\qw\\
&   \qw &\ghost{XX(\pm \frac{\pi}{4})}&\qw&\qw\\
}
\label{EPRCreate}
\end{equation}
The sign of the XX gate depends on the pulse shape solution used for the particular gate, and is stored in a table in the control software, which then determines the appropriate Z rotation to create the Bell state. We use additional rotations to create the other Bell states from this circuit. We measure in the Bell basis by applying a simple CNOT gate followed by a Hadamard gate, and subsequent measurement of the two qubits.
}

\newcommand{\AppendixB}{
\subsection{Appendix B: Implementing and Optimizing Scrambling Operators}

\begin{figure*}
\begin{center}
\includegraphics[width = 160 mm]{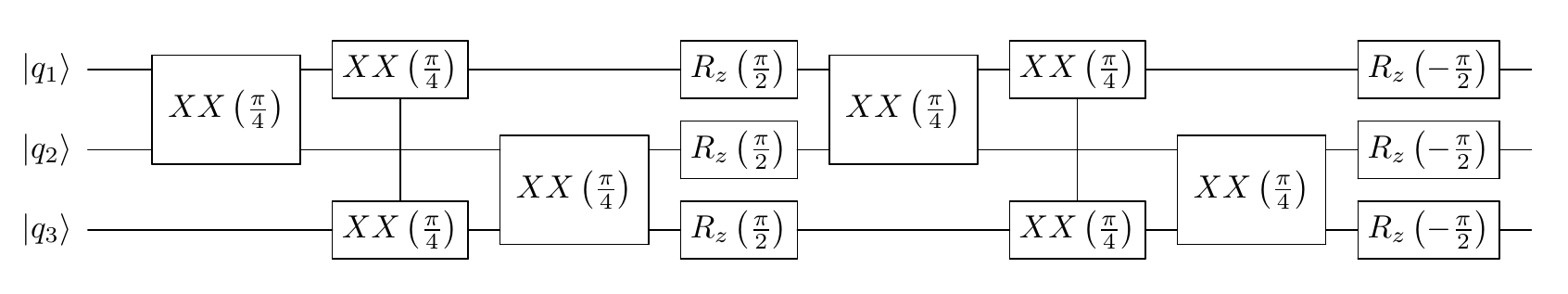}
\end{center}
\caption{Circuit representation of the scrambling unitary used for the probabilistic teleportation scheme (see eqn. \ref{scramblingmatrix}), consisting of six two-qubit entangling XX-gates and individual Z-rotations.}
\label{FigExperimentUXXYY}
\end{figure*}

The scrambling unitary used for the probabilistic teleportation scheme can be represented in the computational basis by the following matrix:
\begin{equation}
\hat{U_s} = \frac{1}{2}\begin{pmatrix*}[r]
-1 & 	0& 	0& 	-1& 	0&	-1& 	-1& 	0\\
0 & 	1& 	-1& 	0& 	-1& 	0& 	0& 	1\\
0 & 	-1& 	1& 	0& 	-1& 	0& 	0& 	1\\
1 & 	0& 	0& 	1& 	0& 	-1& 	-1& 	0\\
0 & 	-1& 	-1& 	0& 	1& 	0& 	0& 	1\\
1 & 	0& 	0& 	-1& 	0& 	1& 	-1& 	0\\
1 & 	0& 	0& 	-1& 	0& 	-1& 	1& 	0\\
0 & 	-1& 	-1& 	0& 	-1& 	0& 	0& 	-1
\end{pmatrix*} 
\label{scramblingmatrix}
\end{equation}
Since this unitary is real, $\hat{U_s}^*=\hat{U_s}$, simplifying the experimental sequence since $\hat{U_d}=\hat{U_s}^* = \hat{U_s}$. The unitary fulfills the following set of equations which verify its scrambling property by showing that it disperses all single-qubit operators into three-qubit operators.
\begin{equation}
\centering
\begin{split}
U^{\dagger} (X\otimes I\otimes I) U = X\otimes Z\otimes Z\\
U^{\dagger} (I\otimes X\otimes I) U = Z\otimes X\otimes Z\\
U^{\dagger} (I\otimes I\otimes X) U = Z\otimes Z\otimes X\\
U^{\dagger} (Y\otimes I\otimes I) U = Y\otimes X\otimes X\\
U^{\dagger} (I\otimes Y\otimes I) U = X\otimes Y\otimes X\\
U^{\dagger} (I\otimes I\otimes Y) U = X\otimes X\otimes Y\\
U^{\dagger} (Z\otimes I\otimes I) U = Z\otimes Y\otimes Y\\
U^{\dagger} (I\otimes Z\otimes I) U = Y\otimes Z\otimes Y\\
U^{\dagger} (I\otimes I\otimes Z) U = Y\otimes Y\otimes Z
\end{split}
\label{scram_equations}
\end{equation}
Here, X,Y,Z, and I are the Pauli operators, and the identity operator, respectively. $\hat{U_s}$ is implemented experimentally using six two-qubit entangling gates (see Fig. \ref{FigExperimentUXXYY}).

In the experimental implementation, we take advantage of the following identity to reduce the number of two-qubit gates needed:
\begin{equation}
     \centering
\centering
\begin{tabular}[c]{l}

\Qcircuit @C=1em @R=0.1em {
&	&\gate{U}&\qw   &&&\qw\\
&\qwx	&&&\push{\rule{.3em}{0em}=\rule{.3em}{0em}}&\qwx\\
&\qwx	&\qw&\qw        &&\qwx&\gate{U^T}\\
}
\label{EPRidentity}
\end{tabular}
\end{equation}
where  the vertical line indicates that the qubits are in an EPR state. 

In this way, we can combine the first two XX gates on ions $2$ and $3$ in $\hat{U_s}$ (and similarly on ions $4$ and $5$ in $\hat{U_d}$) into two single-qubit X-rotations, as shown in Fig. \ref{FigFullExp}.

\begin{figure*}
\begin{center}
\includegraphics[width = 160 mm]{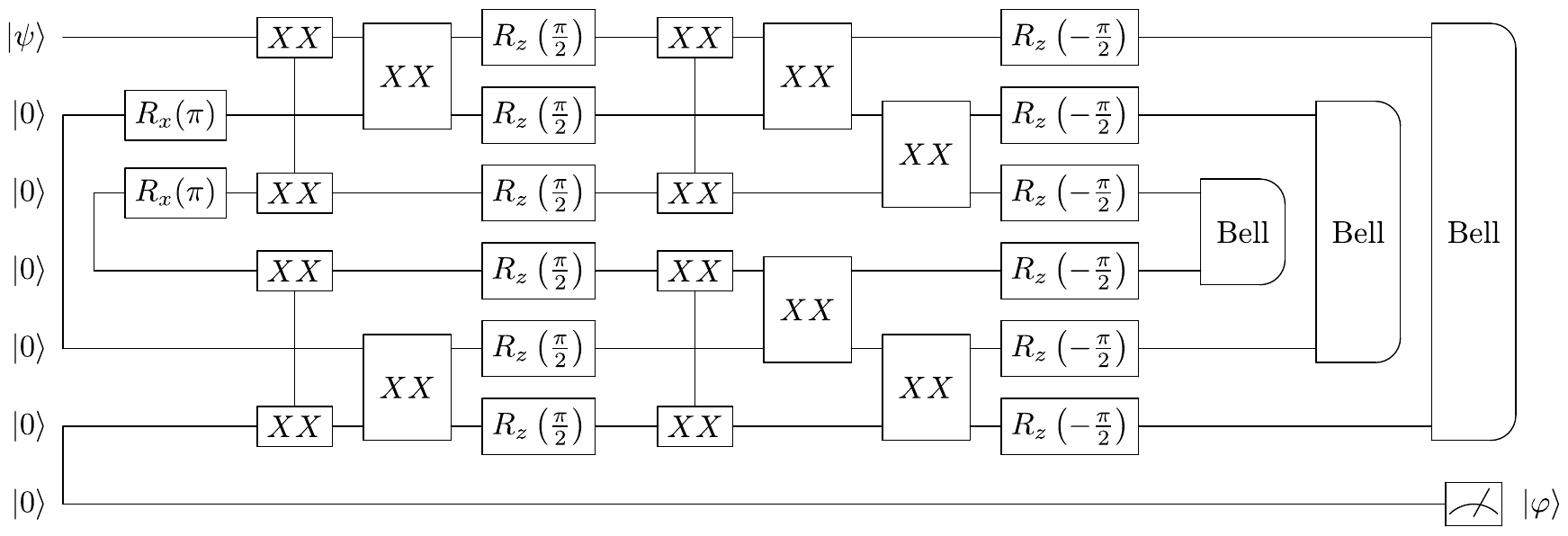}
\end{center}
\caption{The experimental sequence used for the probabilistic teleportation scheme. Any one of the 3 Bell measurements can be used. The scrambling unitary has been simplified using the identity given in eq. \ref{EPRidentity} (see text).}
\label{FigFullExp}
\end{figure*}

In Fig. \ref{ScanScramble} we vary the amount of scrambling in $U$ parametrized by $\alpha$. This is achieved by changing the angles of the Z-rotations depicted in Fig. \ref{FigExperimentUXXYYscan} according to $\theta = \pm \frac{\alpha \pi}{2}$. When $\alpha=0$, the XX-gates combine to create the identity matrix, and when $\alpha=1$, the unitary corresponds to the maximally scrambling case shown in Fig \ref{FigExperimentUXXYY}. Additional Z-rotations are applied around the XX-gates to ensure that $\hat{U_d}=\hat{U_s}^*$. In Fig. \ref{FigScramble}, we measure the delocalization of information throughout the seven-qubit system in the presence of a maximally scrambling unitary $\hat{U}_s(\alpha = 1)$. The Bell measurements used are depicted in Fig. \ref{FigScrambleCircuit} with the same color scheme.

For the data in Fig. \ref{FigOther}, we compose the scrambling unitary depicted in Fig.~\ref{TheoryUCZ}. This unitary has the following matrix representation in the computational basis:
\begin{equation}
U_{CZ} = \frac{1}{2\sqrt{2}}\begin{pmatrix*}[r]
1 & 	1& 	1& 	-1& 	1&	-1& 	-1& 	-1\\
1 & 	-1& 	1& 	1& 	1& 	1& 	-1& 	1\\
1 & 	1& 	-1& 	1& 	1& 	-1& 	1& 	1\\
-1 & 	1& 	1& 	1& 	-1& 	-1& 	-1& 	1\\
1 & 	1& 	1& 	-1& 	-1& 	1& 	1& 	1\\
-1 & 	1& 	-1& 	-1& 	1& 	1& 	-1& 	1\\
-1 & 	-1& 	1& 	-1& 	1& 	-1& 	1& 	1\\
-1 & 	1& 	1& 	1& 	1& 	1& 	1& 	-1
\end{pmatrix*}
\label{UCZmatrix}
\end{equation}
We can confirm that this is indeed a maximally scrambling unitary via a set of equations analogous to Eqn.~\ref{scram_equations}. The optimized circuit used to implement this unitary experimentally is shown in Fig. \ref{FigExperimentUCZ}.  We use a similar circuit to effect the classical scrambling unitary, $\Hat{U_c}$, by implementing only the first three controlled-Z gates. 

Lastly, the Grover search operator labeled $G$ in Fig. \ref{FigOther} is realized using the following circuit:
\begin{equation}
    \centering
\begin{tabular}[c]{l}
\Qcircuit @C=1em @R=0.7em @!R {
&   \multigate{1}{G}	 \\
&   \ghost{G}	\\
}
\end{tabular}
=
\begin{tabular}[c]{l}
\Qcircuit @C=1em @R=0.7em @!R {
&   \qw		&\gate{R_z \left(\pi \right)}	&\gate{R_x \left( \pi \right)}	&\qswap	&\gate{R_z \left( \pi \right)}& \qw \\
&   \qw		&\qw					&\gate{R_x \left( \pi \right)}		&\qswap \qwx	& \qw	& \qw \\
}
\end{tabular}
\label{Gcircuit}
\end{equation}
The SWAP gate (line connecting two X's) is implemented classically by reassigning qubit labels.

\begin{figure*}
\begin{center}
\includegraphics[width = 60 mm]{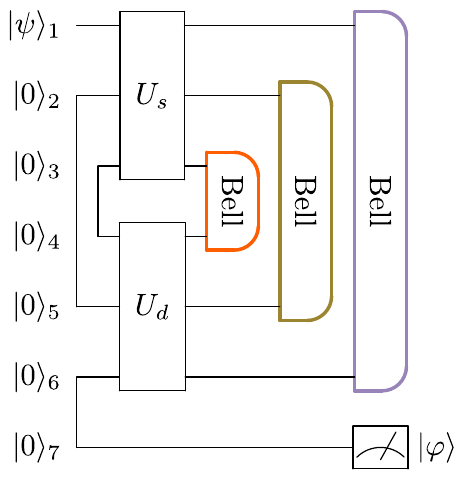}
\end{center}
\caption{Circuit depicting the Bell measurement pairs used in Fig. \ref{FigScramble} (a-b).}
\label{FigScrambleCircuit}
\end{figure*}

\begin{figure*}
\begin{center}
\includegraphics[width = 160 mm]{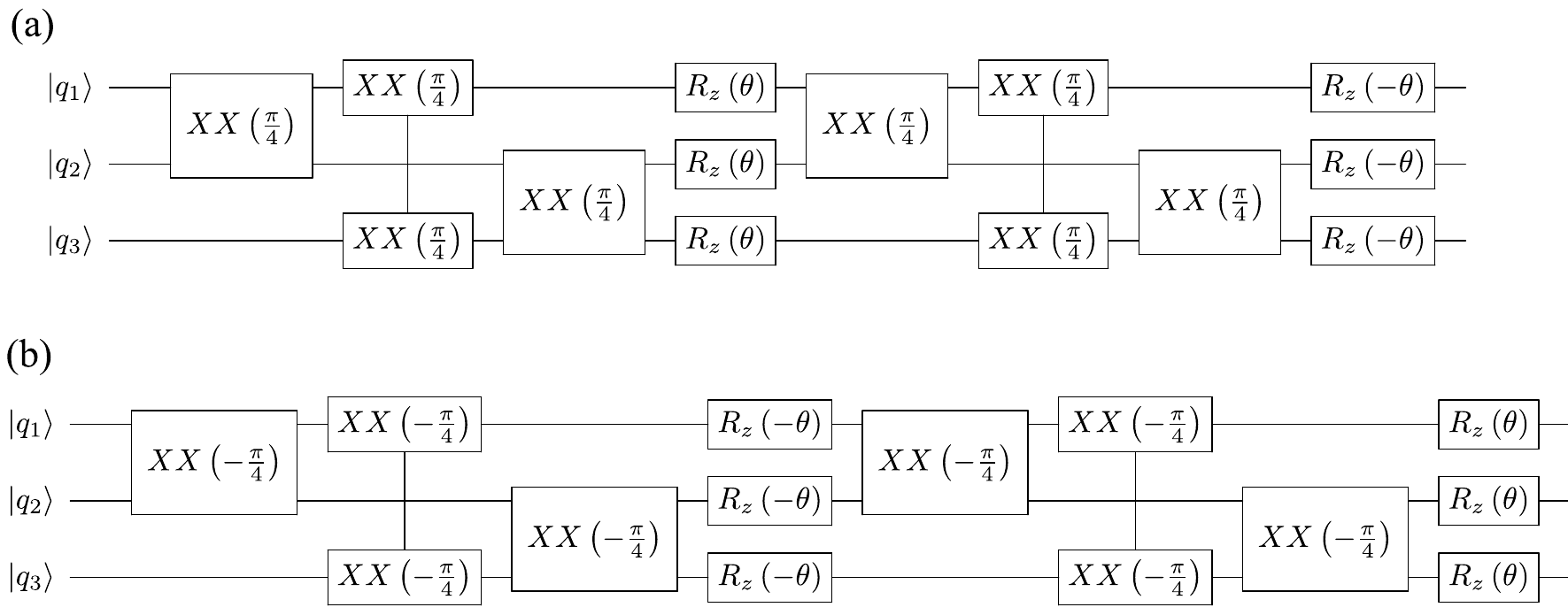}
\end{center}
\caption{(a) Circuit for the unitary used in Fig. \ref{FigScramble} as well as (b) the same unitary with varying degrees of scrambling for the data in Fig. \ref{ScanScramble}. The angles of the the Z-rotations are changed according to $\theta = \pm \frac{\alpha \pi}{2}$ to continuously scan between not scrambling ($\alpha=0$) and maximally scrambling ($\alpha=1$).}
\label{FigExperimentUXXYYscan}
\end{figure*}

\begin{figure*}
\begin{center}
\includegraphics[width = 85 mm]{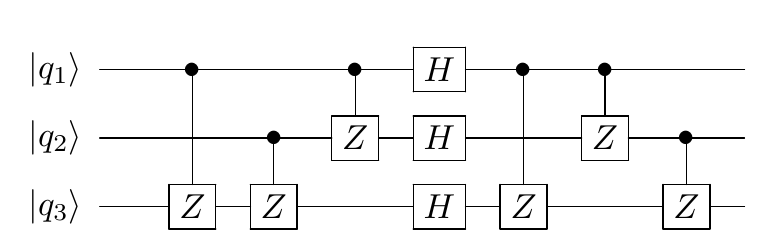}
\end{center}
\caption{Circuit representation of the scrambling unitary from equation \ref{UCZmatrix}, used for the data in Fig. \ref{FigOther}. The breakdown into native gates for the experimental implementation is shown in Fig. \ref{FigExperimentUCZ}. A reduced circuit, made up of only the first three controlled-Z gates, is used to create the classical scrambling unitary $\Hat{U_c}$.}
\label{TheoryUCZ}
\end{figure*}

\begin{figure*}
\begin{center}
\includegraphics[width = 160 mm]{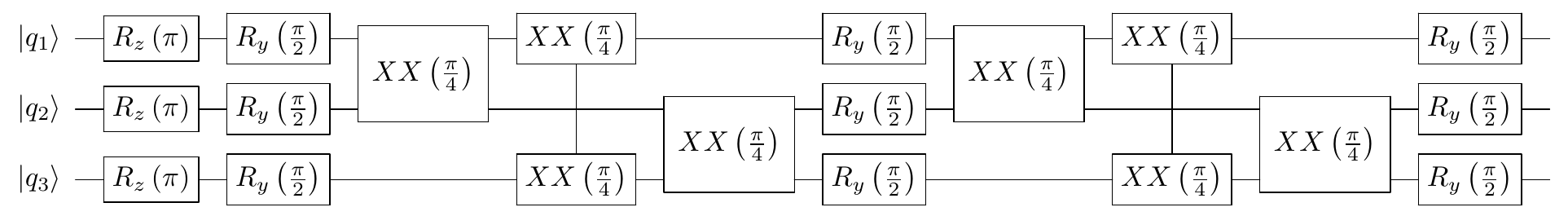}
\end{center}
\caption{The scrambling unitary from equation \ref{UCZmatrix} compiled into native gates. This circuit was used for the measurements in Fig. \ref{FigOther}.}
\label{FigExperimentUCZ}
\end{figure*}

}

\newcommand{\AppendixC}{
\subsection{Appendix C: Numerical Simulations
}
Theory curves were obtained through numerical simulation of the circuits in Appendix B using a simple one-parameter coherent error model. To simulate coherent errors, a random single-(two-)qubit unitary close to the identity was applied following each single-(two-)qubit gate. Single-qubit rotations about the z-axis were performed classically with negligible experimental error and were therefore omitted from this procedure. Random single-(two-)qubit unitary errors were taken as the exponential of a linear combination of the 3 single-(15 two-)qubit traceless Hermitian matrices, with coefficients sampled from a normal distribution with mean 0 and standard deviation $\epsilon/\sqrt{3}$ ($\epsilon/\sqrt{15}$). The resulting observables $\langle P_{\psi} \rangle$ and $\langle F_{\psi} \rangle$ were averaged over $N = 10$ realizations of random error, and the same random errors were used for simulation at each experimental parameter. The error strength $\epsilon = .184$ was chosen to minimize the sum of squared errors in $\langle P_{\psi} \rangle$ and $\langle F_{\psi} \rangle$.
}
\newcommand{\Yb}{$^{171}{\rm{Yb}}^{+} $}
\begin{document}

\title{Verified Quantum Information Scrambling}
\author{K. A. Landsman}\email{kevinlandsman@gmail.com}
\affiliation{Joint Quantum Institute, Department of Physics and Joint Center for Quantum
Information and Computer Science, University of Maryland, College Park, MD 20742}
\author{C. Figgatt}
\affiliation{Joint Quantum Institute, Department of Physics and Joint Center for Quantum
Information and Computer Science, University of Maryland, College Park, MD 20742}
\author{T. Schuster}
\affiliation{Department of Physics, University of California Berkeley, CA 94720}
\author{N. M. Linke} 
\affiliation{Joint Quantum Institute, Department of Physics and Joint Center for Quantum
Information and Computer Science, University of Maryland, College Park, MD 20742}
\author{B. Yoshida}
\affiliation{Perimeter Institute for Theoretical Physics, Waterloo, Ontario N2L2Y5}
\author{N. Y. Yao}
\affiliation{Department of Physics, University of California Berkeley, CA 94720}
\affiliation{Materials Science Division, Lawrence Berkeley National Laboratory, Berkeley, CA 94720}
\author{C. Monroe}
\affiliation{Joint Quantum Institute, Department of Physics and Joint Center for Quantum
Information and Computer Science, University of Maryland, College Park, MD 20742}
\affiliation{IonQ, Inc., College Park, MD 20740}

\begin{abstract}
Quantum scrambling is the dispersal of local information into many-body quantum entanglements and correlations distributed throughout the entire  system. 
This concept underlies the dynamics of thermalization in closed quantum systems, and more recently has emerged as a powerful tool for characterizing chaos in black holes \cite{hayden_black_2007, sekino2008fast, KitaevKITP, Shenker2014, maldacena2016bound}.
However, the direct experimental measurement of quantum scrambling is difficult, owing to the exponential complexity of ergodic many-body entangled states. 
One way to characterize quantum scrambling is to measure an out-of-time-ordered correlation function (OTOC); however, since scrambling leads to their decay, OTOCs do not generally discriminate between quantum scrambling and ordinary decoherence.
Here, we implement a quantum circuit that provides a positive test for the scrambling features of a given unitary process \cite{yoshida_efficient_2017, Yoshida2018}. 
This approach conditionally teleports a quantum state through the circuit, providing an unambiguous litmus test for scrambling while projecting potential circuit errors into an ancillary observable. 
We engineer quantum scrambling processes through a tunable 3-qubit unitary operation as part of a 7-qubit circuit on an ion trap quantum computer. 
Measured teleportation fidelities are typically $\sim80\%$, and enable us to experimentally bound  the scrambling-induced decay of the corresponding OTOC measurement.
\end{abstract}

\maketitle

The dynamics of strongly-interacting quantum systems lead to the memory loss of local initial conditions, analogous to the chaotic behavior of  classical systems \cite{strogatz2018nonlinear}. 
At first glance, this appears inconsistent with the reversible or unitary nature of quantum  time-evolution. 
The resolution lies in the fact that such local quantum information generically becomes entangled with the entire system, and thus hidden in non-local degrees of freedom.
This quantum scrambling process has sharpened our understanding of the limits of thermalization and chaos in quantum systems \cite{hayden_black_2007,sekino2008fast, KitaevKITP, Shenker2014,maldacena2016bound}. 
At one extreme, certain disordered systems can evade thermalization entirely, leading to the slow logarithmic spread of local quantum information \cite{nandkishore2015many}. 
At the other extreme, the existence of a maximum speed limit for thermalization -- known as ``fast-scrambling'' -- is conjectured to occur in certain large-$N$ gauge theories \cite{maldacena_large-n_1999} as well as the dynamics of black holes \cite{hayden_black_2007, sekino2008fast, KitaevKITP,Shenker2014,maldacena2016bound}. 
In particular, scrambling has shed light on the black hole information paradox \cite{Hawking1976,page_average_1993}, suggesting a resolution where quantum information that has crossed the event horizon can indeed be retrieved assuming that the dynamics of the black hole are both  unitary and scrambling \cite{gao2017traversable, maldacena2017diving, SusskindWormhole2017}.

These wide-ranging impacts of quantum scrambling have stimulated the search for experimental evidence \cite{swingle2016measuring,yao2016interferometric,li_measuring_2017, garttner_measuring_2017,meier2017exploring,wei2018exploring,halpern2018quasiprobability} of scrambling dynamics that could help shed light on quantum non-equilibrium processes in exotic materials \cite{blake2017thermal,banerjee2017solvable,ben2018strange} and the fast-scrambling dynamics of black holes \cite{hayden_black_2007, sekino2008fast, KitaevKITP,Shenker2014,maldacena2016bound}. 
Recent work has focused on so-called out-of-time-ordered correlators  (OTOC's) \cite{larkin_quasiclassical_1969,Shenker2014, maldacena2016bound}, which take the form $\langle{\hat{V}^\dagger\hat{W}^\dagger (t)\hat{V}\hat{W}(t)}\rangle$,
where $\hat{V}$ and $\hat{W}$ are  unitary operators acting on separate subsystems. 
The operator $\hat{W}(t) = \hat{U}^\dagger\hat{W}\hat{U}$ is the time-evolved version of $W$ under the unitary operator $\hat{U}=e^{-i\hat{\cal{H}}t}$ generated through either a Hamiltonian  $\cal{H}$ or an equivalent digital quantum circuit sequence.
As scrambling proceeds, $\hat{W}(t)$ becomes increasingly nonlocal, causing the OTOC to decay  \cite{roberts2015localized}, which is taken as an experimental indication of scrambling \cite{li_measuring_2017, garttner_measuring_2017, meier2017exploring,wei2018exploring}. 

However, it is difficult to distinguish between information scrambling and extrinsic decoherence in the OTOC's temporal decay. For example, non-unitary time-evolution arising from depolarization or classical noise processes naturally lead the OTOC to decay, even in the absence of quantum scrambling. 
A similar decay can also originate from even slight mismatches between the purported forward and backwards time-evolution of $\hat{W}(t)$ \cite{garttner_measuring_2017,Yoshida2018, swingle2018resilience}. 
While full quantum tomography can in principle distinguish scrambling from decoherence and experimental noise, this requires a number of measurements that scales exponentially with system size and is thus impractical.

\Figblackhole

In this work, we overcome this challenge and implement a quantum teleporation protocol that robustly distinguishes information scrambling from both decoherence and experimental noise on a family of unitary circuits \cite{yoshida_efficient_2017, Yoshida2018}. 
Using this protocol, we demonstrate verifiable information scrambling and provide a quantitative bound on the amount of scrambling observed in the experiments.

The intuition behind our approach lies in a re-interpretation of the black hole information paradox, depicted schematically in Fig. \ref{Fig1}. 
Let us suppose that an observer (Alice) throws a secret quantum state into a black hole. 
Is it then possible for an outside observer (Bob) to reconstruct this state by collecting the Hawking radiation emitted at a later time? 
Assuming the dynamics of the black hole can be modeled as a random unitary operation $\hat{U}$,  it has been shown that Bob must wait for at least half of the lifetime of the black hole to reliably recover Alice's quantum state \cite{page_average_1993}.
But what if Bob possesses a quantum memory that was previously entangled with the black hole? 
In this case, it was shown that the decoding of Alice's quantum state could be performed by collecting only a few Hawking quanta \cite{hayden_black_2007}.

To this end, an explicit decoding protocol has been recently proposed \cite{yoshida_efficient_2017, Yoshida2018}, which enables Bob to decode Alice's state using only his quantum memory, an  ancillary EPR (Einstein-Podolsky-Rosen) pair, and knowledge of the effective black hole unitary $\hat{U}$ \cite{maldacena2013cool}. 
The protocol  requires Bob to apply $\hat{U}^*$, the complex conjugate of the black hole's evolution operator, to his own quantum memory and one half of the ancillary EPR pair. 
Following this, Bob performs a projective EPR measurement on the emitted Hawking radiation and the corresponding subsystem of his own quantum memory.
This projective measurement plays the role of teleporting Alice's secret quantum state to the reference qubit in Bob's ancillary EPR pair. 
The successful decoding of Alice's quantum information is possible due to the maximally scrambling dynamics of the black hole, which ensure that the information about Alice's secret  state is  distributed, almost immediately,  throughout the entire black hole \cite{hayden_black_2007, hosur2016chaos}.
Thus, the fidelity of quantum teleportation provides a fail-safe diagnostic for true quantum information scrambling. 

Unlike a direct measurement of OTOCs, this protocol can explicitly distinguish scrambling from either decoherence or a mismatch between forward and backward time-evolution (i.e.~the encoding and decoding unitaries: $\hat{U}$ and $\hat{U}^*$). 
Moreover, the success probability of Bob's projective measurement represents an independent measure of the average experimental value of the OTOC, which includes the effects of both noise and decoherence \cite{Yoshida2018}. 
By comparing the teleportation fidelity and the success probability, one can quantitatively and unambiguously bound the amount of quantum scrambling in the unitary operation $\hat{U}$. 

We experimentally implement the above teleportation protocol on a $7-$qubit fully-connected quantum computer \cite{debnath_demonstration_2016} using a family of 3-qubit scrambling unitaries $\hat{U}_s$. 
Our quantum computer is realized with a crystal of trapped atomic \Yb ion qubits, defined by the hyperfine ``clock" states, as described in Appendix A. 
We confine nine ions in the linear ion trap and use the nearly equally-spaced middle seven ions for the circuit. 
We can drive single qubit gates on any of the seven qubits with a typical fidelity of $99.0(5)\%$ and entangling two-qubit gates on any pair of qubits with a typical fidelity of $98.5(5)\%$ (see Appendix A). 
Projective measurements of the qubits in any basis are performed with standard fluorescence techniques \cite{Olmschenk07}, with a qubit readout fidelity of $99.4(1)\%$.  
In combination, EPR state preparation and measurement can be performed with a fidelity of $98(1)\%$ and are generated by a compiler that pieces together native, one- and two-qubit gates to produce the desired gates in a modular fashion \cite{debnath_demonstration_2016}. 

A schematic of the experiment is depicted in Fig.~\ref{Fig1}. The first qubit is prepared in a designated single-qubit state $\ket{\psi}$ and acts as Alice's secret quantum state. We initialize $6$ additional qubits in $3$ non-adjacent EPR pairs, $\ket{\text{EPR}} = \frac{1}{\sqrt{2}}( \ket{00}+\ket{11})$.  Two of these pairs correspond to the black hole entangled with Bob's quantum memory, and the third corresponds to Bob's ancillary EPR pair. We perform a scrambling unitary $\hat{U}_s$ on Alice's qubit and the two black hole qubits, and the decoding unitary $\hat{U}_d = \hat{U}_s^*$ on the two quantum memory qubits and one ancilla qubit. The explicit form of these unitaries and their decompositions into two-qubit gates is detailed in Appendix B. We complete the decoding protocol by projectively measuring any designated pair of qubits -- a chosen Hawking-radiated qubit and Bob's complement of it -- onto an EPR pair. In the absence of decoherence and errors, the probability $P_{\psi}$ of a successful projective measurement is directly related to the OTOC by:
\begin{equation}
P_{\psi} =   \int d\phi \int d\Hat{O}_H \,\,  \left\langle{\Hat{O}_A^\dagger\Hat{O}_H^\dagger (t)\Hat{O}_A\Hat{O}_H(t)}\right\rangle,
\label{OTOCave}
\end{equation}
where $\Hat{O}_A \equiv \ketbra{\psi}{\phi}$ acts on Alice's qubit, $\int d\phi \int d\Hat{O}_H$ denotes an average over single-qubit quantum states $\phi$ and local unitary operators $\Hat{O}_H$ acting on the Hawking-radiated qubit, and $\Hat{O}_H (t) = \hat{U}_s^\dagger\Hat{O}_H\hat{U}_s$.
If the EPR projection is successful, the decoding of Alice's quantum state can be quantified via the teleportation fidelity: $F_{\psi} = |\braket{\varphi}{\psi}|^2$, where $\ket{\varphi}$ is the final state of the ancillary qubit. 
To characterize the nature of different scrambling unitaries, we repeat this protocol for initial states $\ket{\psi} \in \{ \ket{0_x} , \ket{1_x}, \ket{0_y} , \ket{1_y}, \ket{0_z} , \ket{1_z}\}$, where $\ket{0 (1) _{\alpha}}$ denotes the positive (negative) eigenstate of the Pauli operator $\sigma_\alpha$.

\Figscanscramble

We begin by illustrating the challenge associated with interpreting conventional OTOC experimental measurements \cite{li_measuring_2017, garttner_measuring_2017,meier2017exploring,wei2018exploring}. In particular, we perform a control experiment with a non-scrambling unitary  in the presence of deliberate experimental errors (Fig.~\ref{ScanScramble}a): specifically, we take $\hat{U}_s$ to be the identity operation, and introduce single-qubit rotational errors (parameterized by strength $\theta$) following the operation of $\hat{U}_s$, but not the decoding operation $\hat{U}_d$, creating a mismatch between forward and backward time-evolution. 
To allow for a fair comparison with the case of maximally scrambling unitaries, we implement the identity operator as a combination of one- and two- qubit gates of comparable complexity (and total number).
As we increase the size of the mismatch error, we see that the average OTOC (as measured by $P_{\psi}$) decays, consistent with the expected sensitivity of the OTOC to experimental noise (Fig.~\ref{ScanScramble}). Crucially however, the decoding fidelity remains constant near $50\%$, the expected fidelity for an unknown qubit state, confirming that no scrambling has taken place.

Taken together, the measured teleportation fidelity ($F_{\psi}$) and the average OTOC ($P_{\psi}$) enables us to quantify the error-induced decay via an effective noise factor \cite{Yoshida2018},
\begin{equation}
\mathcal{N} \equiv d_A \left [ (d_A +1) \left \langle{}F_{\psi} P_{\psi} \right \rangle - \left\langle{}P_{\psi} \right \rangle \right ],
\label{noise}
\end{equation}
where $d_A$ is the dimension of Alice's quantum state (in our case, $d_A = 2$) and the average is performed over all initial states $\psi$. Note that  $\mathcal{N} =1$ in the ideal case and $\mathcal{N} =0.25$ (e.g.~$1/ d_A^2$) in the fully decohered case.  To this end, the decay of this noise factor from unity indicates the presence of error-induced OTOC decay in our quantum circuit. 
As expected, the observed $\mathcal{N}$ decreases with increasing mismatch (Fig. \ref{ScanScramble}e), reflecting the deliberate error-induced decay of the OTOC, despite the lack of any quantum scrambling dynamics.
Moreover, the measured value of $\mathcal{N} \sim 0.60-0.75$ at zero-mismatch ($\theta=0$) reflects the inherent errors in the experiment, which are expected from the many gates comprising the EPR preparation, unitary operation, and EPR measurement.

With the control experiment in hand, we now characterize information scrambling for a family of unitary operators $\hat{U}_s(\alpha)$ that continuously interpolate (Fig.~\ref{ScanScramble}b) between the identity operator ($\alpha=0$) and a maximally scrambling unitary ($\alpha=1$), as described in Appendix B. The gate decomposition of the $\hat{U}_s(\alpha)$ operation varies only in single-qubit rotations about the $z$-axis, which are performed classically with negligible error. 
Similar to the previous case, we observe the average OTOC to decay as the scrambling parameter, $\alpha$, is tuned from $0$ to $1$, as shown in Fig.~\ref{ScanScramble}c. 
However, unlike the case of the deliberate mismatch-error in Fig.~\ref{ScanScramble}a, the OTOC decay is accompanied by an increase in the decoding teleportation fidelity, indicating the presence of true quantum information scrambling. 
Measurement of a relatively constant noise factor confirms that the experimental error does not depend strongly on the parameter $\alpha$ and thus cannot fully account for the decay of the OTOC. 
In our system, the error scales with the number and type of gates, which are constant across the interpolation.

Using our experimentally measured noise factor $\mathcal{N}$, we can bound the true, scrambling-induced decay of the OTOC for error-free time-evolution via the unitaries $\hat{U}_s(\alpha)$. Assuming that extrinsic decoherence is negligible (i.e.~that coherent errors dominate the experiment), we find that the ideal average OTOC can upper-bounded by \cite{Yoshida2018}:
$4\langle P_\psi\rangle^2/\mathcal{N}^2$.
Therefore, we can experimentally upper-bound the value of the OTOC for the maximally scrambling unitary, $\hat{U}_s(\alpha = 1)$, by approximately 0.47(2).

\Teleportation

In order to demonstrate that our scrambling unitaries are indeed delocalizing information across the entire system, we show that teleportation succeeds independent of the chosen subsystem partition. To do this, we select three different pairs of qubits to be projectively measured in the final decoding step, corresponding to the three qubits acted upon by the scrambling unitary $\hat{U}_s$. 
Decoding succeeds with a fidelity of $70-80\%$ for all projectively measured pairs and all initial states (Fig.~\ref{FigScramble}).  By contrast, the same protocol applied to the non-scrambling identity operator results in a nontrivial decoding fidelity for only a single pair, in which case the entire protocol reduces to the standard setup for quantum teleportation (Fig.~\ref{FigScramble}). Taken together, these measurements demonstrate the full delocalization of Alice's initially local information (the secret qubit state) via the maximally scrambling unitary $\hat{U}_s$. 

Intriguingly, time-evolution that is not maximally scrambling may nevertheless scramble some subset of information. For example, in strongly disordered systems, localization can lead to the scrambling of  phase information but not the scrambling of population \cite{nandkishore2015many}; this situation would correspond to a sort of ``classical scrambler'' wherein teleportation only occurs for $z$-basis states.
 Such classical scramblers may provide insight into  connections between quantum and classical chaos; moreover, measurements of state- and unitary-dependent scrambling may help to more precisely diagnose errors in digital many-body quantum simulations.
To this end, we implement a classical scrambling unitary, $\hat{U}_c$, that completely delocalizes phase, but not population, information: $ [ \hat{U}_c, \sigma_z^i ] = 0$ for all qubits $i$. This commutation implies that no measurement on a black hole qubit can distinguish the initial state $\ket{\psi}$ of Alice's qubit from the phase-reversed state $\sigma_z \ket{\psi}$, so that perfect teleportation is possible only for the classical eigenstates $\ket{0_z}$, $\ket{1_z}$ (see Appendix B for details), as observed in Fig. \ref{FigScramble}.

\Grover

Thus far, our decoding protocols  have all been probabilistic: they rely upon an EPR projective measurement to teleport the unknown quantum state. While the success probability of this EPR projection enables us to quantify the error-induced decay of the OTOC, one can also implement a deterministic version of the decoding protocol \cite{yoshida_efficient_2017, Yoshida2018} (Fig. \ref{FigOther}a). The intuition behind this deterministic decoder is to perform a search for the EPR pair through an implementation of Grover's search algorithm \cite{Grover} instead of post-selecting on a projective measurement. Scrambling remains the focus of this version; while the deterministic decoding fidelity can be lower bounded by 4-point OTOCs, their precise measurement actually corresponds to certain averages of higher-point OTOCs (in our case of a qubit input, 8-point OTOCs).
Such higher-point OTOCs \cite{Shenker2014} can serve as more fine-grained measures of chaos beyond four-point OTOCs \cite{haehl2018fine}, have applications as probes of unitary 4-designs and may be able to diagnose higher-moments of quantum randomness \cite{roberts2017chaos}. Within the Grover-search variant of our decoding protocol, three states are deterministically decoded with an average fidelity of $77(2)\%$ (Fig.~\ref{FigOther}b). 

Finally, we perform a further variation of this protocol that reintroduces a projective measurement as a means to purify errors from the experiment. Here, the same three initial states were decoded with an average fidelity of $78(2)\%$ (Fig.~\ref{FigOther}b). That this purification leads to the same teleportation fidelity despite the additional gate depth (Fig.~4a, purple box) suggests that the fidelity of the  EPR Grover search is roughly $85\%$.

To demonstrate the generality of our approach, these Grover-based protocols were performed with a different class of maximally scrambling unitaries than the previous probabilistic protocol, as described in Appendix B. 

Our work opens the door to a number of intriguing
future directions. 
First, by experimentally scaling to larger circuits, one should be able to probe the scrambling dynamics of Haar random unitaries \cite{emerson_pseudo-random_2003}, complementary to the Clifford scramblers studied here. 
Since the teleportation protocol enables the built-in verification of scrambling, it may provide a natural method for directly measuring the ``randomness''  intrinsic to such circuits, thereby enabling a demonstration of so-called quantum supremacy \cite{boixo2018characterizing}.
Second, while we have focused on the challenges of distinguishing between scrambling and decoherence in this work, our protocol suggests that scrambling circuits may provide a near-term method to benchmark and verify large-scale quantum simulations \cite{cirac2012goals}. 
In particular, by probing the teleportation fidelity as a function of both input state complexity and circuit, one can imagine identifying certain many-body noise mechanisms that would be invisible to typical single- and two-qubit randomized benchmarking methods. 
Finally, it has recently been speculated that quantum simulations can become a natural way to probe the physics of quantum gravity \cite{susskind2017dear}. These arguments hinge on the holographic principle, which suggests that any bulk quantum gravity phenomena can in fact be encoded on the boundary of a quantum mechanical system. To this end, it would be intriguing to modify our teleportation experiment to send a probe qubit into the interior of a black hole before retrieving it via a dynamical back-reaction, thereby providing a way to probe the interior geometry of a black hole. 

Data availability: All relevant data are available from the corresponding author
upon request.

\section{Acknowledgements}
We gratefully acknowledge the insights of and discussions with R. Bousso, D. Harlow, F. Machado, I. Siddiqi, L. Susskind, and Q. Zhuang. Additionally, we thank E. Edwards for the development of Fig. \ref{Fig1}. This work is supported in part by the ARO through the IARPA LogiQ program, the AFOSR MURI on Quantum Measurement and Verification, the ARO MURI on Modular Quantum Circuits, the DOE ASCR Program, and the NSF Physics Frontier Center at JQI. TS and NYY acknowledge support from  the DOE under contract PHCOMPHEP-KA24 and the Office of Advanced Scientific Computing Research, Quantum Algorithm Teams Program. Research at the Perimeter Institute is supported by the Government of Canada through Industry Canada and by the Province of Ontario through the Ministry of Research and Innovation. T.S. acknowledges support from the National Science Foundation Graduate Research Fellowship Program under Grant No. DGE 1752814. Declaration of competing financial interests: C.M. is a founding scientist of IonQ, Inc.

\bibliographystyle{unsrt}
\bibliography{ScramblingBib}

\newpage

\section{Appendices}
\AppendixA
\AppendixB
\AppendixC

\end{document}